# Electronic structure of the substitutional versus interstitial manganese in GaN

Z. S. Popovic[1]*, S. Satpathy[1,2], and W. C. Mitchel[2]

[1]*Department of Physics, University of Missouri, Columbia, MO 65211*

[2]*Air Force Research Laboratory, Materials and Manufacturing Directorate, Wright-Patterson Air Force Base, OH 45433*

Density-functional studies of the electron states in the dilute magnetic semiconductor GaN:Mn reveal major differences for the case of the Mn impurity at the substitutional site ($Mn_{Ga}$) versus the interstitial site ($Mn_I$). The splitting of the two-fold and the three-fold degenerate Mn(d) states in the gap are reversed between the two cases, which is understood in terms of the symmetry-controlled hybridization with the neighboring atoms. In contrast to $Mn_{Ga}$, which acts as a deep acceptor, $Mn_I$ acts as a donor, suggesting the formation of Coulomb-stabilized complexes such as $Mn_{Ga}Mn_I Mn_{Ga}$, where the acceptor level of $Mn_{Ga}$ is passivated by the $Mn_I$ donor. Formation of such passivated clusters might be the reason for the observed low carrier-doping efficiency of Mn in GaN. Even though the Mn states are located well inside the gap, the wave functions are spread far away from the impurity center. This is caused by the hybridization with the nitrogen atoms, which acquire small magnetic moments aligned with the Mn moment. Implications of the differences in the electronic structure for the optical properties are discussed.

Currently there is an intense interest to incorporate magnetic materials into semiconductors for use in spintronics, which seeks to exploit the spin of the electron for novel device applications. The pioneering work of Ohno and coworkers,[1] showing a ferromagnetic Curie temperature as high as 110 K for Mn doped GaAs, demonstrated the feasibility that the ferromagnetic properties can be incorporated into the traditional semiconductors. Since then other dilute magnetic semiconductor systems have been studied. Particular attention has been focused on the Mn doped GaN (GaN:Mn), stimulated in part by the bold theoretical prediction of room-temperature ferromagnetism in the nitrides by Dietl et al.[2] using the Zener model. The recent observation[3] of room temperature ferromagnetism in GaN:Mn, although somewhat controversial, has further accelerated interest in the nitride based dilute magnetic semiconductors.

A knowledge of the electron states introduced by the Mn dopants is an essential ingredient for understanding the microscopic behavior of the system and for tailoring the material properties for potential device applications. The electron states of GaN:Mn have been studied theoretically by a number of authors.[4,5,6,7,8,9,10,11] A main result of these studies is that Mn acts as a deep acceptor in GaN, forming an impurity band detached from the valence band top. This is in sharp contrast to GaAs:Mn, where Mn acts as a shallow acceptor, forming a joint band with the host valence bands, into which holes are introduced. In all these studies, Mn atom was placed at the Ga substitutional site, in line with the traditional wisdom that the 3d impurities are largely substitutional in the III-V semiconductors, while in Si they are largely interstitial.[12] However, there is ample evidence indicating that Mn may occur at sites other than the Ga substitutional site and may even form defect complexes. For example, recent experiments using the Rutherford back scattering have shown an appreciable fraction of the impurities to be interstitial in GaAs:Mn,[13] suggesting that the same may be true for GaN:Mn as well. Similar experiments have not been performed on GaN:Mn to our knowledge; however, samples of Cr doped GaN show only about 90% of Cr to be at the substitutional sites.[14] Thermodynamics arguments suggest that the growth condition could affect the likely sites for the Mn incorporation. For instance, one may be able to stabilize substitutional (interstitial) Mn centers using Ga-poor (Ga-rich) conditions.[15]

Also, there is a major discrepancy between the nominal Mn concentration and the carrier concentration observed in the transport measurements in the entire class of the dilute magnetic semiconductors, where only a small fraction (few percent) of the manganese is found to produce the free carriers. The reason for this low carrier-doping efficiency is thought to be the formation of Mn centers other than the substitutional Mn, including even the complex impurity centers. While the substitutional $Mn_{Ga}$ acts as a deep acceptor in GaN:Mn, an interstitial Mn is expected to act as a donor, as the two Mn(s) electrons of the free atom do not need to participate in covalent bonding. This in turn suggests the formation of the Coulomb stabilized complexes such as $Mn_{Ga}$-$Mn_I$-$Mn_{Ga}$. Formation of other types of complexes is also likely and in fact clusters such as $Mn_xN_y$ have been suggested to occur in GaN:Mn.[16,17] It is therefore clear that before an adequate understanding of the ferromagnetic behavior in the dilute magnetic semiconductors can be developed, it is imperative to understand the nature of the Mn impurity centers in the host material. In this paper, we study the electron states of the substitutional versus the interstitial Mn using density-functional methods.

Our calculations were performed for the wurtzite crystal structure using the supercells $Ga_{15}Mn_{Ga}N_{16}$ and $Ga_{16}Mn_IN_{16}$ for the cases of substitutional and interstitial Mn, respectively, which corresponds to about 6% Mn concentration. We used the local spin-density approximation (LSDA) to the density-functional theory as implemented in the linear muffin-tin orbitals (LMTO) method in the atomic spheres approximation.[18] Christensen's empirical approach[19,20] was used to correct for the well known problem of the band-gap underestimate in the semiconductors; however, we found that this correction does not change any essential physics of the problem. The bulk lattice constant for GaN was used and also any local relaxations around the Mn atom were not considered. This is reasonable since, as was shown recently for the substitutional $Mn_{Ga}$ in GaN, local relaxations have negligible effect[4] on the electronic and the magnetic properties and the same may be expected to be true for the interstitial $Mn_I$ as well.[21]

Fig. 1 shows the calculated band structure for the wurtzite $Ga_{0.94}Mn_{0.06}N$, with the Mn dopants at the Ga site. Like GaAs:Mn, the material is half-metallic, with the Mn(d) states split into a triply degenerate $t_2$ state, which occurs at the Fermi energy, and a doubly degenerate $e$ state just above the top of the valence bands.

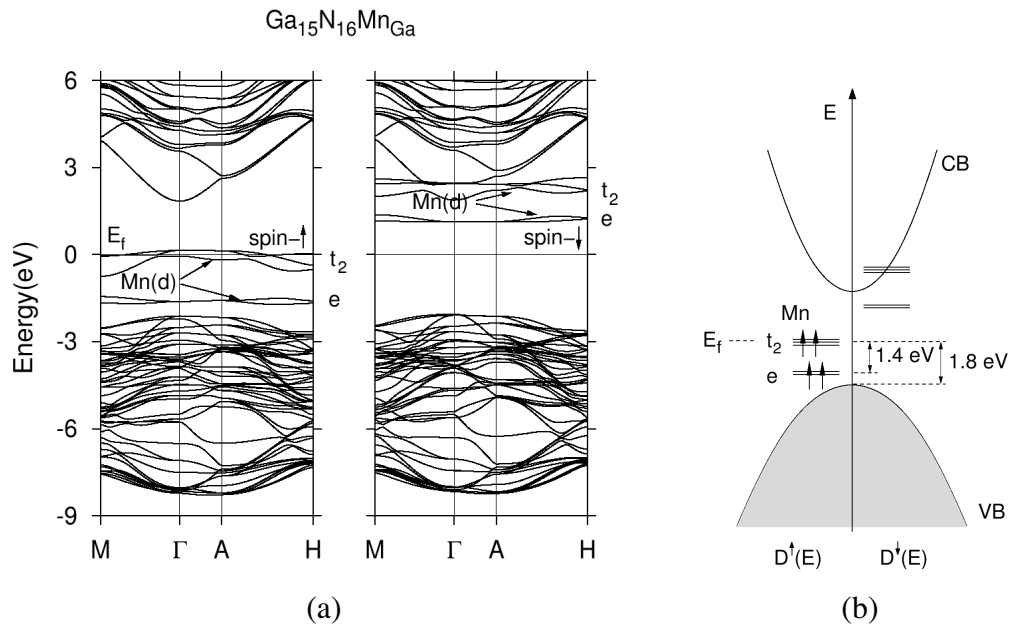

(a)                    (b)

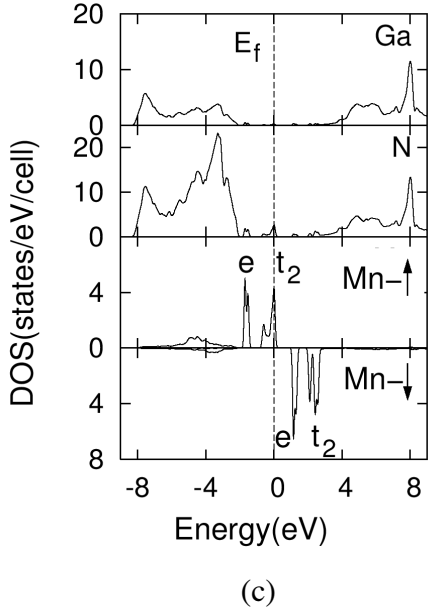

[Fig. 1. Density-functional electronic structure of the ferromagnetic $Ga_{15}MnN_{16}$ with Mn at the substitutional Ga site. Shown are the majority and minority spin bands (a), the schematic positions of the Mn levels (b), and the atom resolved densities-of-states (c).]

(c)

Concerning the position of the e state, there is a controversy in the literature, viz., whether the *e* state is in the gap[9] or inside the valence bands.[4] We find that the difference is due to a subtle feature, viz., whether the Ga(3d) electrons are treated as valence or core electrons. The reason why the position of the Mn(e) state might be so sensitive to small changes in the potential becomes clear by considering the symmetry properties of the various orbitals. In essence, as follows from the discussion below, a strong anti-bonding interaction with the N(p)-like valence bands ensures the $Mn(t_2)$ state to lie above the valence bands, while such interaction is absent for the Mn(e) level, making its position insensitive with respect to the energy of the N(p) bands.

The $Mn_{Ga}$ ($Mn_I$) has approximate tetrahedral (octahedral) symmetry and it is convenient to use these symmetries for the Mn(d) states. For the case of the substitutional Mn, $Mn_{Ga}$ is surrounded by four nitrogen atoms in a tetrahedral surrounding and ignoring the further neighbors has the tetrahedral symmetry. The $Mn_{Ga}$(d) orbitals span the $e+t_2$ representations of the tetrahedral $T_d$ group, while the surrounding $N("p_z")$ orbitals span the $a_1+t_2$ representations, as seen from the symmetry properties summarized in Table I. The $N("p_z")$ orbitals refer to a local atom-based coordinate system and, by definition, point towards the adjacent Mn atom. The strong pd interaction, symmetry-allowed for $t_2$ but not for e (as there is no e component for $N("p_z")$), produces a N-character $t_2$ bonding state in the valence band and a Mn-character $t_2$ anti-bonding state, raising its energy. The Mn(e) state is allowed to interact only with the remaining $N(p_x/p_y)$ orbitals, which, although they do span the *e* representation, interact only weakly with Mn because their lobes do not point towards the Mn atom. This explains why the Mn($t_2$) state occurs above the Mn(e) state for Mn$_{Ga}$. This also explains why the occupied states at the Fermi energy, which occurs in the middle of the Mn($t_2$) bands, should have the Mn(d)-N($p_z$) antibonding character, as is readily visible in the charge-density contour plot (Fig. 2), where the charge-density is zero at the middle of the Mn-N bond.

Table I. Irreducible representations spanned by the atomic orbitals in the $T_d$ and the $O_h$ groups, relevant for Mn$_{Ga}$ and Mn$_I$ for the wurtzite structure. The symbol N("$p_z$") denotes the nitrogen p orbitals with lobes pointed towards the Mn atom at the center of the tetrahedron or the octahedron.

| Point Group | Orbitals | Number of orbitals | Irreducible Representations |
|---|---|---|---|
| $T_d$ (tetrahedral) | Mn(d) | 5 | $e+t_2$ |
| | N(s) or N("$p_z$") | 4 | $a_1+t_2$ |
| | N(p) | 12 | $a_1+e+t_1+2t_2$ |
| $O_h$ (octahedral) | Mn(d) | 5 | $e_g+t_{2g}$ |
| | N(s) or Ga(s) or N("$p_z$") or Ga("$p_z$") | 6 | $a_{1g}+e_g+t_{1u}$ |
| | N(p) or Ga(p) | 18 | $a_{1g}+e_g+t_{1g}+t_{2g}+2t_{1u}+t_{2u}$ |
| $C_{6v}$ | Mn(d) | 5 | $a_1+e_1+e_2$ |

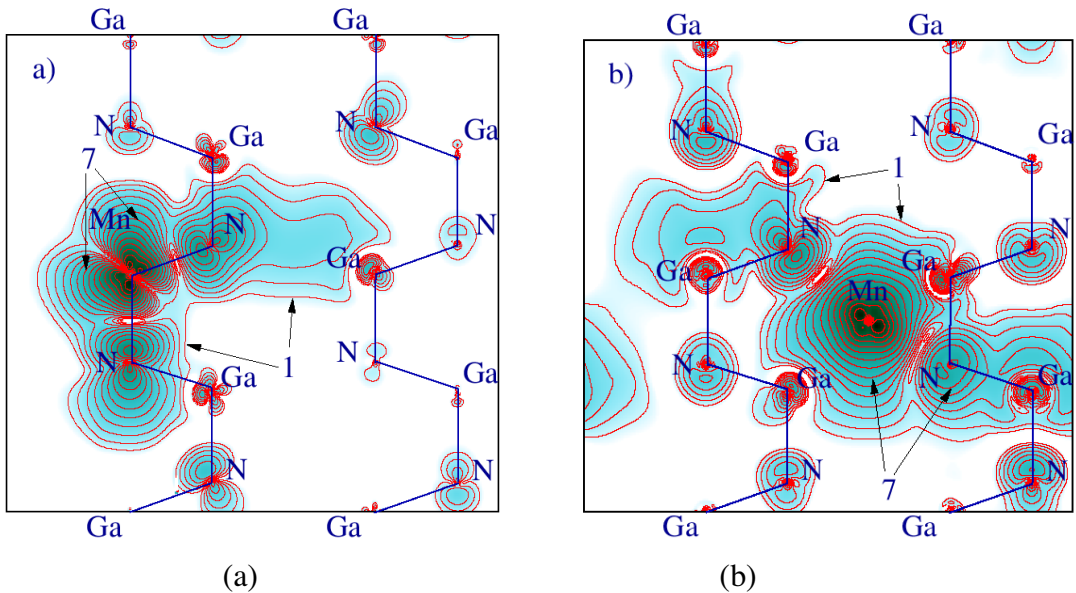

(a)          (b)

[Fig. 2 (color online): Charge-density contours corresponding to the occupied bands near $E_F$ for Ga(Mn)N: substitutional Mn$_{Ga}$ (a) versus interstitial Mn$_I$ (b). Size of the energy windows were 1.0 eV and 1.2 eV, respectively, so as to include all states of the highest occupied Mn band near $E_F$. Note, from the zeros of the charge density at the Mn-N bond center, the anti-bonding character of the Mn-N hybridization and the orientation of the N("$p_z$") orbitals (local z axis points to Mn). Note also the spread of the charge to distant nitrogen atoms. Of the twelve atoms that form the two octahedra around $Mn_I$ (Fig. 3), only four are visible in (b) (the four closest to Mn). We used the supercells, $Ga_{31}Mn_{Ga}N_{32}$ and $Ga_{32}Mn_IN_{32}$, i.e., with about 3% Mn, to compute the charge densities and the contour values are: $\rho_n = \rho_0 \, 10^{n\delta}$, where $\rho_0 = 2.1 \, x \, 10^{-3} \, e \, / \, \mathring{A}^3$, $\delta = 0.25$, and $n$ labels the contours.]

According to the density-functional results, $Mn_{Ga}$ is a deep acceptor with energy of about 1.8 eV, in excellent agreement with the acceptor energy extracted from the optical absorption measurements.[22] For the dipole-allowed optical transitions, the matrix element of the position operator between the initial and the final electron states $< \Psi_i \mid \vec{r} \mid \Psi_j >$ is non-zero, where $\vec{r}$ is along the direction of polarization of light. For transitions involving the Mn states, we may just consider the atoms surrounding the Mn, viz., the four N atoms forming the tetrahedron and just the N(p) orbitals, which have the major contribution to the valence and the conduction bands. For a strong transition involving the Mn atom and the host bands, the N("$p_z$") orbitals must be involved; other nitrogen p orbitals would have weaker matrix elements. Considering just the $MnN_4$ cluster with the tetrahedral $T_d$ symmetry, one finds that all transitions among the Mn e and $t_2$ levels, the valence bands (VB), and the conduction bands (CB) are dipole allowed. This leads to the following features in the optical absorption: (a) A sharp Mn (e) $\rightarrow$ Mn ($t_2$) transition at 1.4 eV, (ii) A VB $\rightarrow$ Mn ($t_2$) absorption band starting at 1.8 eV, (iii) A VB $\rightarrow$ Mn (e) absorption band starting at 3.3 eV, and (iv) The VB $\rightarrow$ CB absorption starting at the gap value of 3.5 eV, all of which are consistent with the observed optical absorption.[22]

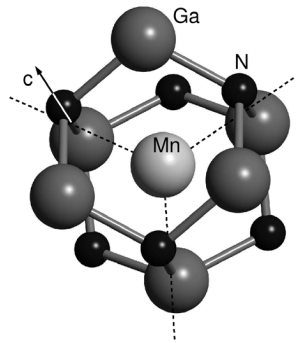

[Fig. 3. The interstitial position in the wurtzite structure with approximate octahedral symmetry, with the N and the Ga atoms forming two separate octahedra surrounding the Mn.]

We now turn to the case of the interstitial Mn, which we place at the largest void in the wurtzite structure occurring at the center of the triangle joining the midpoints of the three adjacent Ga-N bonds pointed along the *c* direction (Fig. 3). In view of the near-octahedral symmetry about the void (two interpenetrating octahedra of six Ga or six N atoms, whose centers are slightly offset with respect to each other and with respect to the position of the void), it is instructive to examine the symmetry properties of the electron states in terms of the full octahedral group $O_h$.

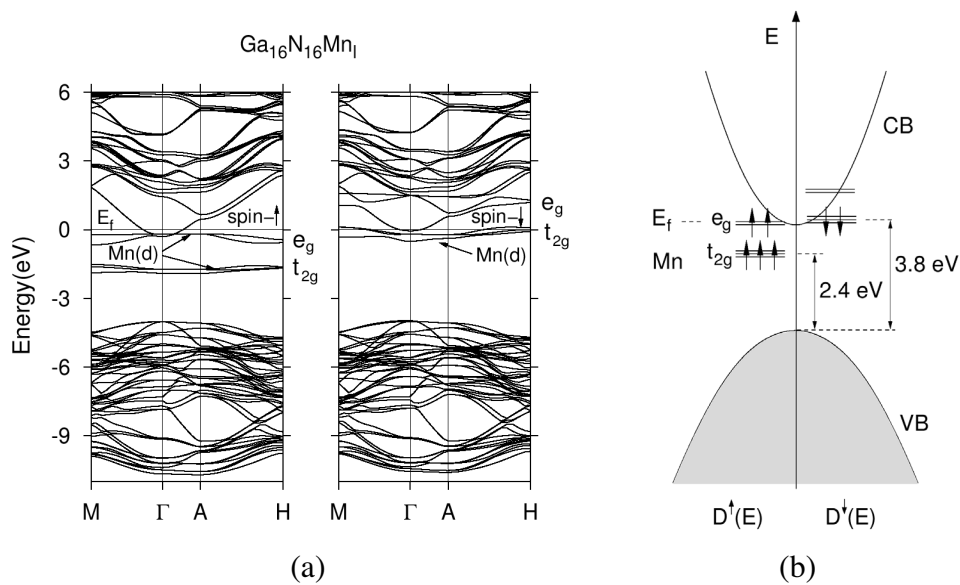

(a)

[Fig. 4. Density-functional band structure for the interstitial $Mn_I$ in GaN (a) and the band diagram extracted from it (b).]

The electronic structure for the interstitial $Mn_I$ is shown in Fig. 4. We find the interstitial manganese to have the low-spin configuration Mn ($d^5\uparrow$, $d^2\downarrow$) as compared to the high-spin ($d^5\uparrow$, $s^2$) configuration for the free atom, which may be rationalized by the increase of the energy of the manganese s states by the particle-in-a-box confinement by the twelve surrounding Ga/N atoms. We find the magnetic moment to be 2.6 $\mu_B$ for $Mn_I$ as compared to 3.5 $\mu_B$ for $Mn_{Ga}$. In both cases, the adjacent nitrogen atoms acquire a small magnetic moment of about 0.05 $\mu_B$ aligned along the Mn moment.

Unlike the substitutional case, the three-fold degenerate state ($t_{2g}$) occurs below the two-fold degenerate eg state, a result that can be understood following the same line of symmetry argument we used for the substitutional case. As mentioned earlier, although the Mn position does not have the exact octahedral symmetry (Mn-N distances for the surrounding six nitrogens are for example different, but only by $\sim 0.4$ Å, 2.1 Å vs. 2.5 Å), an approximate octahedral symmetry does exist, which splits the Mn (d) states into an $e_g + t_{2g}$ combination of the $O_h$ group. The N ("$p_z$") states, which are expected to interact the most with the Mn orbitals on account of their orientation towards the Mn atom, span only the $a_{1g} + e_g + t_{1u}$ symmetry. Of significance here is the fact that a $t_{2g}$ component is absent. Therefore the interaction with the occupied N ("$p_z$") states in the valence band does not affect the Mn ($t_{2g}$) states (symmetry-forbidden), while it pushes up the Mn ($e_g$) state in energy, which in the process acquires an anti-bonding Mn-N character. Exactly the same type of symmetry considerations hold for the Ga ("$p_z$") orbitals as well. This explains the relative positions of the $t_{2g}$ and the $e_g$ states. The analysis is supported by the figure showing the charge-density contours (Fig. 2b), where the anti-bonding interaction with the N/Ga ("$p_z$") states is clearly seen from the zeros of the charge-density at the Mn-N and Mn-Ga bond centers. Thus, to summarize these arguments, in the $Mn_{Ga}$ case with tetrahedral symmetry, since the occupied nitrogen-"$p_z$" orbitals also have the $t_2$ component, they push the Mn ($t_2$) state up with respect to the Mn (e) state, while in the case of $Mn_I$ with octahedral symmetry, the Mn ($e_g$) state gets pushed above Mn ($t_{2g}$) following the same logic.

The dipole allowed optical transitions may be examined as before by considering the $MnN_6Ga_6$ cluster (Fig. 3) and the $O_h$ symmetry. Unlike the substitutional case, no sharp lines are predicted in the optical spectra, as the $t_{2g} \rightarrow e_g$ transition is not dipole allowed. Other transitions lead to broad bands and are easily masked by the substitutional $Mn_{Ga}$ atoms. However, there would be a clear difference between the two cases in the photoluminescence spectra. Also because of the difference in the Mn(d) local densities-of-states in the two cases, the Mn X-ray spectra should also be different, as has been suggested for GaAs:Mn.[5]

As indicated from the band structures, the interstitial $Mn_I$ acts as a donor, while the substitutional $Mn_{Ga}$ acts as a deep acceptor. This suggests the formation of manganese complexes consisting of, for example, one interstitial and two substitutional manganese atoms, where the two extra electrons on $Mn_I$ neutralize the holes on the $Mn_{Ga}$ atoms, thus resulting in a neutral triad with passivated carriers. The Coulomb energy becomes particularly favorable for such a structure. We have computed the electronic structure of such a complex ($Mn_{Ga}Mn_IMn_{Ga}$). The results confirm the picture of carrier passivation, with a $d^5$ configuration for all three atoms. We find the two $Mn_{Ga}$ to be ferromagnetically aligned opposite to the moment of the interstitial $Mn_I$, leading to a net ferromagnetic moment.

Acknowledgments

We thank John Albrecht for stimulating this work and acknowledge support of this work by the U. S. Department of Energy through Grant No. DE-FG02-00ER45818. SS would also like to thank the National Research Council and the Air Force Office of Scientific Research for support of this work through a Summer Faculty Fellowship in the Air Force Research Laboratory, Dayton, Ohio.